# Synchronous PIV measurements of a self-powered blood turbine and pump couple for right ventricle support

Kagan Ucak[1] Faruk Karatas[1] Emre Cetinkaya[2] Kerem Pekkan[1]

[1]Mechanical Engineering, Koç University, Rumelifeneri, Istanbul, 34450, Turkey
[2]Mechanical Engineering, Yildiz Technical University, Istanbul, Turkey

**Address for Correspondence:**

Kerem Pekkan, PhD.

Professor

Mechanical Engineering Department Koç University

Rumeli Feneri Kampüsü, Sarıyer, Istanbul, Turkey

Phone: +90 (533) 356 3595

Fax:+90 (212) 338 1548

E-mail: kpekkan@ku.edu.tr

## Abstract

A blood turbine-pump system (iATVA), resembling a turbocharger was proposed as a mechanical right-heart assist device without external drive power. In this study, the iATVA system is investigated with particular emphasis on the blood turbine flow dynamics. A time-resolved 2D particle image velocimetry (PIV) set-up equipped with a beam splitter and two high speed cameras, allowed simultaneous recordings from both the turbine and pump impellers at 7 different phased-locked instances. The iATVA prototype is 3D printed using an optically clear resin following our earlier PIV protocols. Results showed that magnetically coupled impellers operated synchronously. As the turbine flow rate increased from 1.6 to 2.4 LPM, the rotational speed and relative inlet flow angle increase from 630 to 900 rpm, and 38 to 55% respectively. At the trailing edges, backflow region spanned 3/5 of the total passage outlet flow, and an extra leakage flow was observed at the leading edge. For this initial turbine design, approximately, 75% of the turbine blade passage was not contributing to the impulse operation mode. The maximum non-wall shear rate was ~2288 s-1 near to the inlet exit, which is significantly lower than the commercial blood pumps, encouraging further research and blood experiments of this novel concept. Experimental results will improve the hydrodynamic design of the turbine impeller and volute regions and will be useful in computational fluid dynamics validation studies of similar passive devices.

**Keywords**: Blood turbine, blood pump, PIV, ventricular assist device, RVAD, VAD, right ventricular, right heart failure, iATVA.

# 1 Introduction

There are approximately 26 million patients worldwide suffering severe heart failure (Bowen et al. 2020). Due to the donor shortage, only a handful of patients ever receive cardiac transplantation, resulting in 1.6 million deaths annually (Tsao et al. 2023). Long-term mechanical circulatory support (MCS) with an artificial heart is a solution, but external power needs result in driveline infections which is a major cause of device failure (Blume et al. 2016). State of the art MCS also require power supplies and control units significantly limiting the patient mobility and result poor quality of life (Dunn et al. 2019). For patients with right-heart failure, the current MCS therapies are palliative and result uncontrolled gradual worsening of the cardiovascular performance due to remodeling (Gorter et al. 2018). Finally, currently a dedicated right-heart assist device is not clinically available.

State-of-the-art MCS devices are also complex and produced through advanced manufacturing techniques where the device costs typically reach €120,000 per patient, excluding the hospital charges. Thus, these expensive systems are not affordable for many underserved patients and are not accessible globally. As such, devices are limited to few standardized sizes and could not be customized for the individual patient needs or changing disease conditions. To address all these grand challenges and eliminate device complications an integrated aortic-turbine venous-assist system (iATVA) is proposed by our group (Pekkan et al. 2018). In its most basic configuration, the novel blood turbine of iATVA harvests hydrodynamic energy from the aorta and transfers it to the right-ventricle enabling a self-contained fully-implantable right-ventricular support. Its first proof-of-concept is demonstrated in single-ventricle disease pediatric circulation where high venous pressures are reduced significantly, with low aortic steal and almost no effect on the blood oxygen saturation level (Pekkan et al. 2018).

Despite its encouraging potential, as a truly novel MCS, the detailed hemodynamics and operational efficiency of iATVA is largely unknown. Particularly, we hypothesized that the blood damage budget of the turbine section would be very high as it operates at inefficient high-pressure and low-flow design specific speeds. Therefore, to have a better understanding of the operational hemodynamics of the iATVA, in this study a detailed PIV measurement campaign is undertaken focusing its first original design version. A detailed turbine velocity field analysis of iATVA is reported here with contribution of simultaneously measured pump velocity fields. While there are several excellent PIV studies focusing on traditional ventricle assist device (VAD) pump components (Shu et al. 2015; Rowlands et al. 2018), blood turbine hemodynamics is novel and emerges as a unique contribution to the literature. As such understanding the working efficiency and defining operating conditions of iATVA will initiate design changes and hopefully enable improvements in its performance.

In a typical device design workflow, hemodynamic performances and efficiencies of blood pumps have been analyzed via computational fluid dynamic (CFD) simulations (L. Wiegmann et al. 2017; L. Wiegmann et al. 2019; Gil et al. 2023). While CFD is a practical approach, simulations of ventricle assist devices always require experimental validation since high variations in the computational results are typical (Hariharan et al. 2018; Ponnaluri et al. 2023). According to its turbine specific speed, the optimal iATVA design requires an impulse turbine regime which has a very limited validation dataset. Furthermore, the synchronous simulation of coupled pump and the turbine sections is also technically challenging. Finally, expected jet-flow regimes of iATVA turbine introduces additional requirements for the CFD solvers. Therefore, PIV experiments as presented here are justified to document the actual performance and operational characteristics of the iATVA system prior to CFD.

Expanding on our recent experimental PIV study focusing on the US Food and Drug Administration (FDA) benchmark blood pump (Ucak et al. 2024), in this paper we acquired entire impeller and volute velocity fields of the iATVA turbine using a traditional time-resolved PIV system. A novel beam splitter and a mirror are utilized to divide the laser into both turbine and pump sides, which enabled the synchronous accusation of PIV results from both impellers, first-time in literature. The manuscript is organized as follows; Following the methodology section, the presented results include checking of the magnetic coupling via simultaneous turbine and pump side images. After the check, we compared three operating conditions for both impeller sides. Later, we compared three different turbine sections to extend our investigation in turbine flow dynamics. To obtain quantitative comparisons, we compared shear rate, jet flow, and backflow magnitudes for different operating conditions and impeller phases. Finally, we suggested fluid dynamic design improvements for the future iATVA versions.

## 2. Materials and Methods

### 2.1. 3D printed transparent iATVA for PIV

The original iATVA design, reported in (Pekkan et al. 2018) which incorporates 14 neodymium magnets for coupling turbine and pump impellers, was revised for PIV experiments (see Fig. 1). A protocol for manufacturing optically transparent 3D printed PIV pump prototypes was already developed in our recent work on FDA benchmark blood pump (Ucak et al. 2024) which is also applied to the present iATVA system. Briefly, the casings of the turbine and pump sides were altered with flat surfaces to eliminate the light reflections. Clear Resin V4 was used to print the models using a stereolithography printer (Form 3, Formlabs, Massachusetts, US) at a layer thickness of 25 microns, and the iATVA parts were then cured. The difference between the diameter sizes of 3D printed parts and the computer model was less than 0.15 mm. The refractive index of the transparent parts was measured as 1.52 via a gem refractometer. The parts of the iATVA system were manufactured by +90 Rapid Prototyping (Istanbul, Turkey).

## 2.2. Experiment set-up

The setup included a beam splitter (50R/50T Standard Plate Beamsplitter, Edmund Optics, NJ) and a mirror to illuminate both sides of iATVA simultaneously (see Fig. 2). Two high speed cameras (Phantom Miro M310, Vision Research Inc, NJ) were synchronized with a 5W continuous double pulse Nd-YAG laser (LaVision GmbH, Germany). An external pump rotated the turbine, and inlet flow rate adjusted with a calibrated rotameter. The outlet of the turbine directed the flow to the reservoir that was located lower than the outlet to minimize the head pressure. 1 L of distilled water was used at the room temperature of 24° C degrees. The pump had its own loop with maximum 69 mm vertical height relative to the center of pump inlet. 13 μm sized fluorescent polymer particles (Fluoro-Max 36- 3B; Thermo Scientific Inc., Fremont, CA) were dissolved homogenously for the turbine water reservoir and pump loop.

Phase-locked images were obtained via the magnets inside the impellers and the hall effect rotational-speed sensor which created signals under high magnetic field changes. Those signals were controlled via an Arduino microprocessor to capture 7 phases of both turbine and pump sides. Time differences between image pairs for each orientation were selected as 90 μs.

## 2.3. Operating conditions

Three different conditions (C1, C2, C3) were investigated synchronically in both turbine and pump sides according to the flow rate of the turbine as 1.6, 2.0 and 2.4 LPM. For each measurement, seven different impeller phases were recorded on both turbine and pump sides with a total of 2660 phase-locked double frame pairs. This has led 190 double frame pairs were obtained for each phase impeller phase. Each measurement was repeated three times to evaluate the experimental uncertainty. Initially, measurements were completed in the location of the bottom, middle and top plane of the turbine impeller for C2. Then, other conditions were measured only for the middle plane due to the large amount of experimental data.

### 2.4. PIV processing

Images were calibrated with a calibration plate and manually scaled with impeller diameters. Pre-processing was applied with subtracting background images with 8 pixel sizes. Multi pass algorithm with decreasing size of 48x48 and 24x24 pixel was performed. Post processing of smoother and outlier eliminators was applied for eliminating specious vectors. The relative velocity fields were obtained by subtracting the rotation of the impellers in MATLAB via in house scripts provided in the Supplementary Material 2.

### 2.5. Verification

PIV post-processing convergence test is performed for phase locked images which showed 1% convergence error after 60 number of image pairs for both the turbine and pump sides. Still for the best practice, typically 190 images were utilized in the velocity field processing (Section 3.2-3.6). Time-resolved velocity fields presented in this article (Section 3.1) employed 5 consecutive images of the turbine and pump sides with 3 and 5% convergence, respectively (Supplementary Material 1). The PIV set-up was validated in our previous study using a transparent FDA benchmark blood pump manufactured identically, which agreed by 93.7% correlation with the PIV experiments performed by Hariharan et al. (Ucak et al. 2024).

### 2.6. Hemodynamic performance parameters

Simultaneous analysis was performed with instantaneous (averaging 5 sequence) velocity fields to investigate magnetic coupling of the turbine and pump impellers. Other analyses were performed by averaging 190 sequences. Standard deviation was calculated from the averaged datasets in between repetitions of three measurements as our previous study (Ucak et al. 2024). Reynolds number was defined as $Re = \frac{\rho n D^2}{\mu}$ where $\rho$ represents density, $n$ is impeller rotation speed, $D$ is the diameter of the rotor, and $\mu$ is dynamic viscosity. The flow coefficient, $\varphi = \frac{Q}{nD^3}$ was calculated where $Q$ is the flow rate. The absolute velocity values along suction, inter-blade and pressure lines were investigated according to chord line of the turbine impeller.

Additionally, to assess the blood damage, shear strain was calculated in five locations in the volute and as displayed in Fig. 7 according to the equation of $\dot{\gamma} = \frac{du}{dy}$ where $\dot{\gamma}$ is shear rate, and $u$ is velocity component orthogonal to the $y$ direction.

## 3. Results

### 3.1. Synchronous velocity fields of pump and turbine impellers

Instantaneous velocity fields of the pump and the turbine impeller are acquired at the same time instance (Fig. 3). In these measurements, distinct flow recirculation zones are observed along the turbine impeller. Initially, at the impeller phase 1 (P1), no vortices are detected. Then at phase 2 (P2), in which the blade leading edge is aligned with the turbine inlet nozzle entrance, three distinct vortical structures are shed. Two of these vortices rotate opposite to the turbine rotation and one rotate in the same direction. At the next phase (P4), the jet-like turbine inlet flow coming from the nozzle hits the next blade passage in row. This event converted two of the vortices to the downstream. In addition, a backflow occurred just after the next blade passage swept the inlet jet. This process indicates a close association with backflow and the jet inlet flow in the turbine. At phase P6, most of the blade passage is free of major circulations and follows a similar flow structure with phase P1. Thus, the vortex structures are observed in ~60% of each cycle. To orient the readers; the blade aligned with the inlet nozzle is labeled as blade 2 (B2), and the downside neighbor blade of B2 is labeled as B1 in Fig 3. The flow mostly moves along the pressure side of the B1. The backflow coming from the pressure side of B2 is observed in the suction side of B2, and it moved until the jet flow in the volute. Interestingly, when the backflow is highest, it covers 3/8 of total blade passage flow between B1 and B2.

For the pump side insignificant variations were observed among different phases. The flow velocity increased towards the volute side, and the maximum velocity was significantly lower compared to the turbine. Recirculation zones were observed at P1, P2, P4 and P6 time points. In Supplementary Material 1, we tracked tip locations of phase locked turbine and pump blades and compared their rotations for all frames. Magnitudes of the rotation of both magnetically coupled impellers were the same, thus we confirmed that the pump and turbine worked simultaneously.

In the turbine, the maximum velocity measured relative to the impeller were 1.8, 2.1, 1.3, 1.7 m/s for different phases of P1, P2, P4, P6 consecutively. They occurred around the trailing edge of the pressure sides except the P2 with occurrence around the leading edge of the pressure side. For the pump side, the maximum velocity occurred near to the volute side for each phase. This section provided temporal characteristics of magnetically coupled impellers, employing only a small number of PIV frames (see Section 2.6). The detailed quantitative comparison providing accurate velocity magnitudes are presented in the following sections, averaged over 190 image pair sequences.

### 3.2 Flow patterns and effect of operating conditions

During the altered flow conditions of C1, C2 and C3, rotation speed of iATVA is measured as 630, 790, and 900 rpm, respectively. Between the three repetitions of experiments performed in these conditions, the standard deviation of the rotation speed is maximum 1% for each impeller. For the turbine, Re numbers in impeller region for different conditions range between 100.000 and 150.000, whereas for the pump, Re numbers are between 75.000 and 100.000. Interestingly, flow coefficients are found to be 0.0075 with less than 2% difference in between different operating conditions. Similar flow patterns are observed in the middle plane of the turbine and pump among all conditions as presented in Fig 4. The backflow passing over trailing edge of B2 creates a strong recirculation zone for each case, and it is mostly conserved in all conditions. For the pump side, recirculation still exists at the same location for all conditions and flow patterns do not change significantly. Therefore, we conclude that flow patterns in the turbine and pump do not change dramatically with the flow rate of the turbine around the flow rates studied.

The maximum relative velocities are obtained in the jet flow just on the pressure side of B1. The magnitudes of these peak velocities are 1.45, 1.75, 1.90 m/s for C1, C2, C3, and they change almost linearly with the flow rate.

### 3.3 Turbine flow at different sectional planes

Flow patterns at different sectional heights influence the overall blood turbine performance. Measurements acquired at three different PIV laser sheet heights are displayed in Fig 5 and blade leading and trailing edge heights are 5.1 and 6.1 mm. Overall as we move from the blade bottom to the impeller shroud, increased flow movements are recorded. Around the trailing edge of the Blade 2, labeled as B2 in Fig. 5, a backflow is observed in all planes. In the lowest plane, the flow coming from the inlet is visible in between B2 and B3, as there exists a strong inlet jet flow on the pressure side of Blade 1 (B1). Apart from the flow fields induced by the turbine inlet jet, most of the blade passage regions are relatively still and do not critically contribute to the outlet flow. For the middle measurement plane, jet inlet flow is also significant. The contribution of the middle plane to the outlet flow is monitored to be more than the bottom plane. At the plane closest to the shroud, the inlet jet flow is more dominant on blade B2 pressure side. Likewise, the suction side of blade B2 reveals that the flow passes on top of the blade around the leading edge. This leakage flow encounters the inlet jet flow around the leading edge and causes a change of the inlet jet flow direction to shroud side instead of to the center of the rotor.

The maximum relative velocities are obtained with 2.0 and 2.1 m/s magnitude around the trailing edge of the pressure side for the bottom and middle planes. At the top plane, the maximum velocity magnitude reaches 1.3 m/s near to the inlet side.

### 3.4 Turbine impulse analysis

Effects of the jet flow and backflow are further quantified for all three conditions at two impeller orientations as displayed in Fig. 6. This quantification is important to better analyze how blade passage regions contribute to the turbine function and which locations are crucial for these contributions. Velocity profiles display how most of the jet flow is localized around the pressure side of blade B1. In the orientation of P1, 60% of the outlet flow in this blade passage region is maintained only by area spanning 25% of the region, which corresponds to the jet flow 1(J1). This ratio is interestingly conserved in all three conditions. J1 region is thinner at the leading edge of blade B1, then expanded towards to the center of the turbine. The ratio of jet flow at P2 (J2) to J1 is increasing as 39, 45, 55% for conditions of C1, C2, and C3, respectively. This increment indicates that the load on the second blade passage region increases with the flow rate. In the orientation of P2, a strong backflow (BF) starts at the trailing edge of B2, and it diffuses until the nozzle side. The magnitude of J1 linearly increases with the flowrate, but the magnitude of the BF does not change. Thus, ratio of the BF to J2 decreases as 63, 54 and 42% for conditions of C1, C2, and C3 respectively.

### 3.5 Turbine volute and nozzle flow

The absolute velocity field in the turbine and its nozzle are displayed in Fig. 7 for three different conditions. Also, the velocity profiles along pressure, inter-blade and suction lines are plotted for all cases. The flow characteristics are mostly similar in all conditions but increments in the velocity magnitudes at the volute region, pressure sides of the blades, nozzle region, and impeller tip are observed. For C2, the pressure side displays almost one-time higher magnitude than the suction and inter-blade lines. All three profiles follow a small increment of velocity magnitude along the blade direction except the region close to the leading edge (after 12 mm). At the leading edge, magnitudes in the inter-blade line suddenly increase with 1260 $s^{-1}$ incline, and relatively smaller inclines are observed in the pressure (750 $s^{-1}$) and suction (280 $s^{-1}$) lines. Pressure velocity profiles of C3 and C1 display 22.6% higher and 18.6% lower average velocity difference with C2 which agree with their flow rate differences. For the inter-blade line, C3 and C1 display ~27.5% higher and ~18.4% lower velocity magnitudes than C2 in each location. After 12 mm, similar inclines of C1 (1319 $s^{-1}$) and C3 (1000 $s^{-1}$) are observed with C2. Interestingly, the minimum average difference is measured in the suction side in between different conditions. C3 exhibits almost the same velocity profile as C2, and the average velocity magnitude of C1 is 21.1% less than C2. Furthermore, we investigated the different phases of C2 to improve the understanding of the transition of the blade tips from the inlet entrance.

The result of the inlet passage phases and corresponding velocity magnitudes along suction, inter-blade and pressure lines are displayed in Fig. 8. Significant variations in the velocity profiles are observed just after the 12 mm location. Upstream of this location, lines along the pressure, inter-blade, and suction regions are similar. Inter-blade velocity profile displays the most diversity in between sequential frames, and the slope of the inclement is linearly correlated with the distance in between the inlet entrance and the inter-blade line tip. Increments at the time 0, 0.011, and 0.022 s are 190, 290 and 610 $s^{-1}$, respectively. At the suction line, increments are 6, 70 and 180 $s^{-1}$ according to the same time order. Furthermore, the largest magnitudes of the increments are measured at the pressure side with the values of 635, 835 and 945 $s^{-1}$ for the same time order.

### 3.6 Shear rate measurements

To assess and compare expected blood damage, instantaneous velocity gradients and the resulting shear force experienced by the blood cells are critical. Calculated shear rates by the impeller speed are as 855, 1065, and 1222 $s^{-1}$ for C1, C2, and C3 respectively. Shear rates of three conditions calculated from the velocity field, are presented in Table 1 for five locations with the mean of the corresponding standard deviations. The shear rate increases from C1 to C2 for all locations except L5. L1 displays the least shear rate for all conditions among other locations. The shear rate decreases from inter-blade location to pressure side (from Location 2 to 4). C3 displays mostly similar results to C1 but it has a very high magnitude of standard deviation in the volute region. Thus, it does not obey the linear trend of other conditions. L5 is in the blade passage (near to the leading edge and pressure side), and C1 and C3 have relatively close results like other locations. Yet, C2 displayed a smaller shear rate at L5 than other conditions.

## 4. Discussion

Due to their potential to impact underserved communities globally, medical devices based on *frugal science* gained a lot of interest (Sarkar and Mateus 2022; Valiathan 2018; Corsini, Dammicco, and Moultrie 2021). Still the application of frugal concepts to Class III blood-wetted devices has been challenged due to their technical complexity and stringent regulatory compliance requirements. Passive devices, like the present iATVA concept, with no electrical components, allows fully customized patient-specific 3D printed blood pumps that can reach inaccessible low socio-economical patient populations. This approach can address the current extremely low market penetration (~1% globally) of artificial mechanical assist devices through reduced the device costs up to 15 times. The iATVA concept bears an exciting competitive advantage over the existing clinically approved devices (Atti et al. 2022; Gil et al. 2023) and offer an alternative pathway to achieve the long-sought use of intracorporeal VADs as a destination therapy.

Present *in vitro* study provided an important insight on the blood damage budget of the turbine section of iATVA. It is well established that high blood flow shear values directly correlate with higher hemolysis in VADs (L. Wiegmann et al. 2017). For example, the red blood cell damage is reduced when the shear rate gets low ($<300$ $s^{-1}$), furthermore, Hashimoto considered maximum shear rate of 1000 $s^{-1}$ in their in vitro analysis, indicating their disregard for examining damage over even higher shear rates (Hashimoto 1989). Tamari et. al. mentioned that BP-50 centrifugal pump peak shear rate is 9200 $s^{-1}$ for 3600 rpm (Tamari et al. 1993). Bourqe et. al. used a detailed CFD analysis to observe the hemocompatibility of the Heartmate III and the shear rate did not exceed 18518 $s^{-1}$ for around 6000 rpm (Bourque et al. 2016). Wang considered an axial heart pump and implemented both constant and pulsatile flow, the maximum shear rate is calculated as 14280 $s^{-1}$ at 7000 rpm and this value covers 3-5 % of the volume (Wang et al. 2020). The measured shear rate values of iATVA, for both turbine and pump sections, are considerably lower than these clinically approved devices. For example, iATVA's peak shear rate values varied between 386 $s^{-1}$ to 2288 $s^{-1}$. These low shear rates are attributed to the small size and low rotational speed of iATVA (790 rpm), which is sufficient as the right-heart assist does not require very high pump pressure heads and the required aortic flow steal is low. In most regions the shear rates generated in iATVA turbine is lower that the physiological flow shear rates encountered in a healthy person's arteries, which is less than 1740 $s^{-1}$ (Bourque et al. 2016). Likewise, for cats Lipowsky et. al. stated maximum physiological shear rate as 2000 $s^{-1}$ (Lipowsky, Usami, and Chien 1980). An improved hydrodynamic design of iATVA would easily achieve shear rates lower than 2000 $s^{-1}$.

Results of this study suggested several design improvements for the impeller, shroud, and nozzle of the iATVA turbine in terms of hemodynamic performance. In the previous work Increasing the number of blades in the turbine was considered (Pekkan et al. 2018), which is also supported by the current PIV experiments. Usage of the splitters, or partial blades, is preferred over increased blade numbers due to their lower surface area which is not desired due to increased thrombogenicity. Also, splitters would not block the turbine outlet excessively. Likewise, a significant flow swirl is observed at the turbine exit, requiring stationary outlet guide vanes at the outlet pipe. Sang and Zhou stated that proper outlet guide vane angle can regulate the flow and eliminate undesired large scale vortical structures (Sang and Zhou 2017).

Another hydrodynamic improvement can target the observed inlet flow separation. Due to the relatively high velocity jet inflow flow, the blade passages do not contribute equally to the power extraction. Furthermore, this causes non uniformity in shear rates inside the volute region. Thus, instead of a single inlet volute, usage of multiple inlets by dividing the total inlet flow would increase the hydrodynamic performance and the balance of the blade passages. Despite this backflow, the second blade passage still able to contribute to the outlet flow significantly. Therefore, three equally spaced inlets for the six blades will increase the efficiency pending further design studies. Present experimental results will be useful to validate a computational fluid dynamics model which can be used in systematic design optimization studies.

Current iATVA design requires tight tolerances due to its small inlet jet wake size relative to the leading edge blade height. Flow field measurements showed that only half of the impulse jet flow is contained and captured by the turbine leading edge. In addition, the leakage gaps between the turbine blades and the shroud are large (0.4 and 0.75 mm at the trailing and leading edges, respectively) to control blood damage due to high shear between the shroud and the rotating impeller. As a remedy the blade height at the inlet can be increased so that it matches the jet wake, which will allow us to utilize the full thrust coming from the volute. However, the inequality in leakage gaps at the edges causes undesired blade passage flow, so the shroud should be designed parallel to the blade height. At the trailing edge the blood damage due to flow shear is less critical since the impeller speeds are lower. Therefore, secondary flows from the pressure side to suction side can be reduced through decreased thickness of the leakage gap and partially-closed impeller design.

Finally, due to some unintended experimental challenges this study had some limitations. First, the usage of the opaque impellers limited the full impeller velocity field capture due to shadows. Still different impeller orientations are studied to overcome this limitation. As such, at the pump side, to achieve a smooth inflow, a long inlet pipe was obstructing the camera view. The transparent iATVA prototype used in this study had very thin walls and relatively 2D impellers, thus we ignored the refractive index mismatch effects. This does not affect the results in the blade passages and most of the shroud region but influenced velocity profiles in the nozzle leading higher but conservative estimates of the shear rates reported in these regions.

## 5. Conclusion

This study provided a detailed hydrodynamic understanding on the first design version of the iATVA system. The simultaneous operation of the turbine and pump sides was analyzed through a novel beam-splitter PIV set-up, with emphasis on the blood turbine section, as it can also be used as a standalone implanted device. Presented experimental results are critical for further performance improvement and refinement studies. Our analysis identified four critical design improvements. First, the usage of the small splitters in between the turbine blades will avoid the observed recirculating regions and achieve a more parallel relative flow through the impeller. Next, the excessive gap at the trailing edge reduces turbine power therefore a partially closed impeller can be beneficial. Likewise, outlet impeller blade height can be increased to capture a larger section of the turbine inlet jet. Finally, the flow speed through the nozzle inlet exceeds 5 m/s which is too high for this narrow section. Thus, a turbine with multiple inlets can be considered to reduce possible blood damage. As such, a variable gap between the shroud and the impellers can be considered as a tradeoff between low leakage flow and high shear rates.

The iATVA concept is ultimately intended to be used as a fully implanted cardiovascular device. Therefore, the desire to achieve a compact device and anatomical locations of inlet/outlet grafts constrained the use of a centrifugal turbine at the impulse regime. Despite the reduced efficiency, the hydrodynamic performance of a centrifugal turbine operating close to the impulse mode is found to be encouraging. Still there is a significant room for further improvements in turbine hydrodynamics leading more efficient cardiovascular energy extraction through turbine and delivery to the pump side. Eventually, with further refinement, we envision this technology will fully support pulmonary blood flow in patients with single ventricle defects, offering promising prospects and enhanced patient care. While the current analysis provided valuable insights, further investigations are needed to validate the new turbine design versions.

## Acknowledgements

Funding was provided by European Research Council (ERC) Proof of Concept *BloodTurbine*, and TUBITAK BİDEB - 2247A - 120C139.

# List of Tables:

**Table 1**. iATVA turbine shear rate comparison for three operating conditions. Locations are displayed in Fig. 7.

| Condition | L1 (s$^{-1}$) | L2 (s$^{-1}$) | L3 (s$^{-1}$) | L4 (s$^{-1}$) | L5 (s$^{-1}$) | Std (%) |
|---|---|---|---|---|---|---|
| **C1** | 386 | 1866 | 1756 | 1486 | 760 | 3.92 |
| **C2** | 436 | 2288 | 2016 | 1718 | 655 | 8.06 |
| **C3** | 462 | 1817 | 1825 | 1525 | 819 | 11.75 |

## List of Figures:

**Fig. 1** Assembly drawing of the iATVA system and its components. A circular array of magnets couples the turbine and pump impellers. The turbine and pump employ dedicated bearings and housings with complete separation of arterial and venous blood in the device. The trailing and leading edges of the turbine and pump blades are marked in the figure together with the respective rotation directions.

**Fig. 2** Selected views of the experiment set-up are provided. Two high speed cameras (HSC 1 and HSC 2), a hall-effect sensor-based trigger system controlled by an Arduino processor and the beam splitter are labeled. Other flow loop components include rotameter, reservoir, external pump, and power source for the external pump. Red and blue colors are used to highlight arterial and venous sections of iATVA, respectively.

**Fig. 3** The simultaneous measurements of turbine and pump velocity fields at the second operating condition (Condition C2) are displayed. In total, seven phases were recorded for both impeller sides. However, since phases were mostly similar, only four of the phases are displayed for brevity. Averaging was employed with 5 phase-locked velocity fields to resolve the transient operation.

**Fig. 4** Relative velocity fields in turbine and pump impellers for three operating conditions (Conditions C1, C2 and C3). Operating conditions are determined according to the turbine flow rate ranging from 1.6 to 2.4 LPM. Turbine blades are labeled the same as in Fig. 3. Same color bar is used with Fig. 3 and for both turbine and pump sides. The mean jet flow velocity magnitudes in the pressure sides of B1 are 1.2, 1.5, 1.9 m/s for C1, C2, and C3 respectively.

**Fig. 5** Velocity fields which are relative to the impeller rotation on the turbine impeller at three different sections (bottom, middle, and top) for the operating condition of turbine flow at 2 LPM. Raw images are presented in the first row and the corresponding velocity fields are presented in the second row. The blade aligned with the nozzle entrance is labeled as blade 2 (B2), and the next blade is labeled as B1. The leakage flow that passes through upon the gap in between the blade and the shroud is captured in the top section. The high speed of the jet flow dominates the velocity field, so the color bar is limited to 0.75 m/s to emphasize the flow structure in the blade passages. The mean jet flow velocity magnitudes at the pressure side of B1 are 1.3, 1.5, and 1.1. for each sections.

**Fig. 6** Quantitative visualization of velocity fields at two phases with velocity profiles. **a)** Most of the jet flow (J1) goes to the first blade passage in the left turbine phase, and the phase where the jet flow (J2) moves to the second blade passage is displayed in the right side. Backflow (BF) is marked in the right velocity field. **b)** The ratio of the velocity magnitudes of J2 to J1 for different operation conditions is displayed on the left side with an increasing trend. The ratio of the mean velocity magnitude of BF to J2 is presented on the right side with a decreasing trend. The mean BF velocity magnitudes were constant for different operation conditions.

**Fig. 7** Velocity profiles along three different lines that are parallel to blade curvature. a) Absolute velocity contours for three different operating conditions. The suction, inter-blade and pressure lines are displayed in the first column. b) Velocity profiles along different sections and operating conditions. Radius is defined from the trailing edge to the leading edge (R). The top left plot is the comparison of velocity fields for operating condition C2. Other plots are comparisons of the sections with three different operating conditions, C1, C2 and C3.

**Fig. 8** Velocity profiles along three different lines for three different phases (P1, P2, P3). Phases are displayed in the left side figure. Suction, inter-blade and pressure lines are displayed on the right side for each phases. Radius (R) is defined the same as in Fig. 7. Pressure sections had missing points near to the beginning of the radius due to the shadows, thus they were eliminated from the dataset.

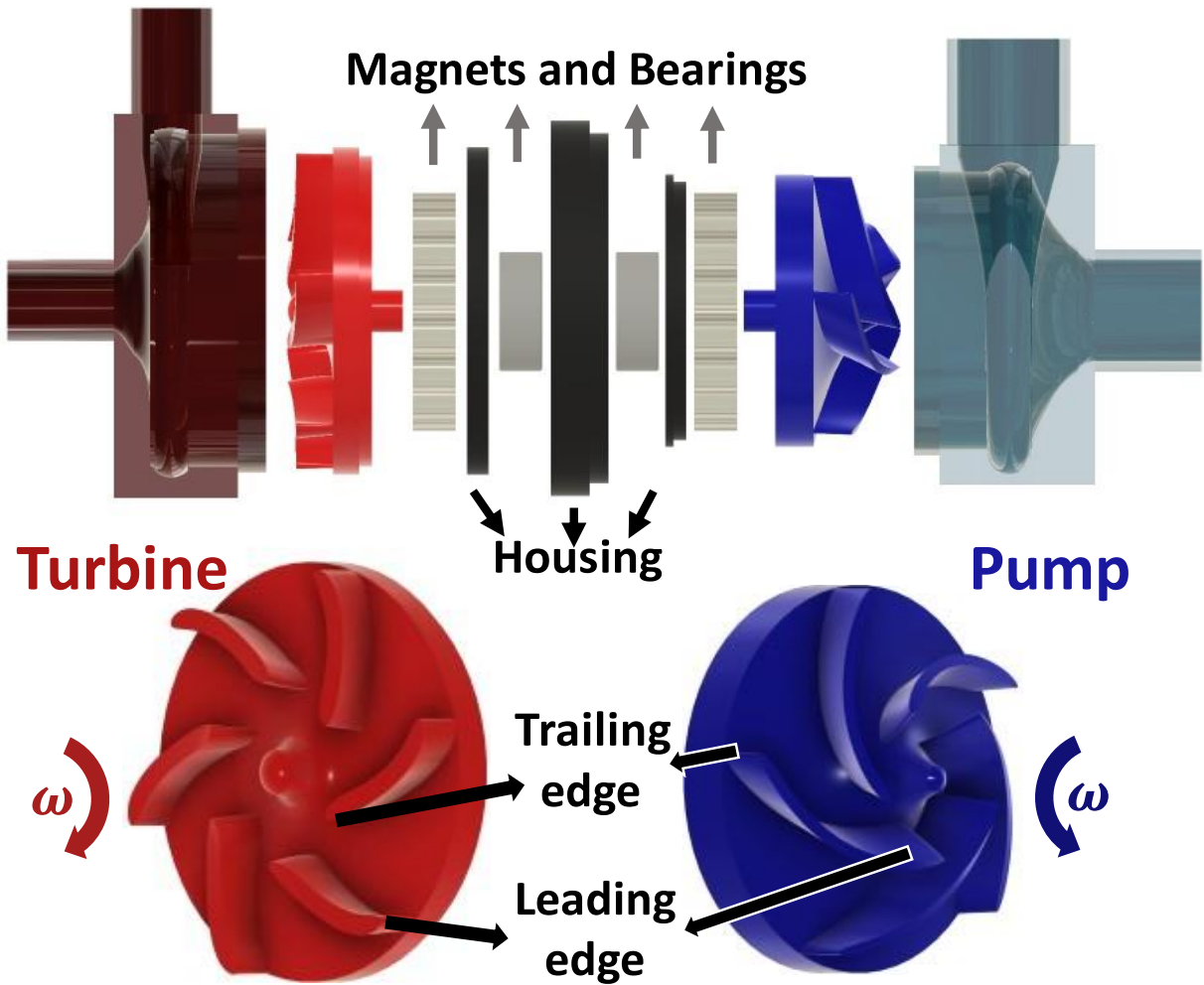

**Fig. 1** Assembly drawing of the iATVA system and its components. A circular array of magnets couples the turbine and pump impellers. The turbine and pump employ dedicated bearings and housings with complete separation of arterial and venous blood in the device. The trailing and leading edges of the turbine and pump blades are marked in the figure together with the respective rotation directions.

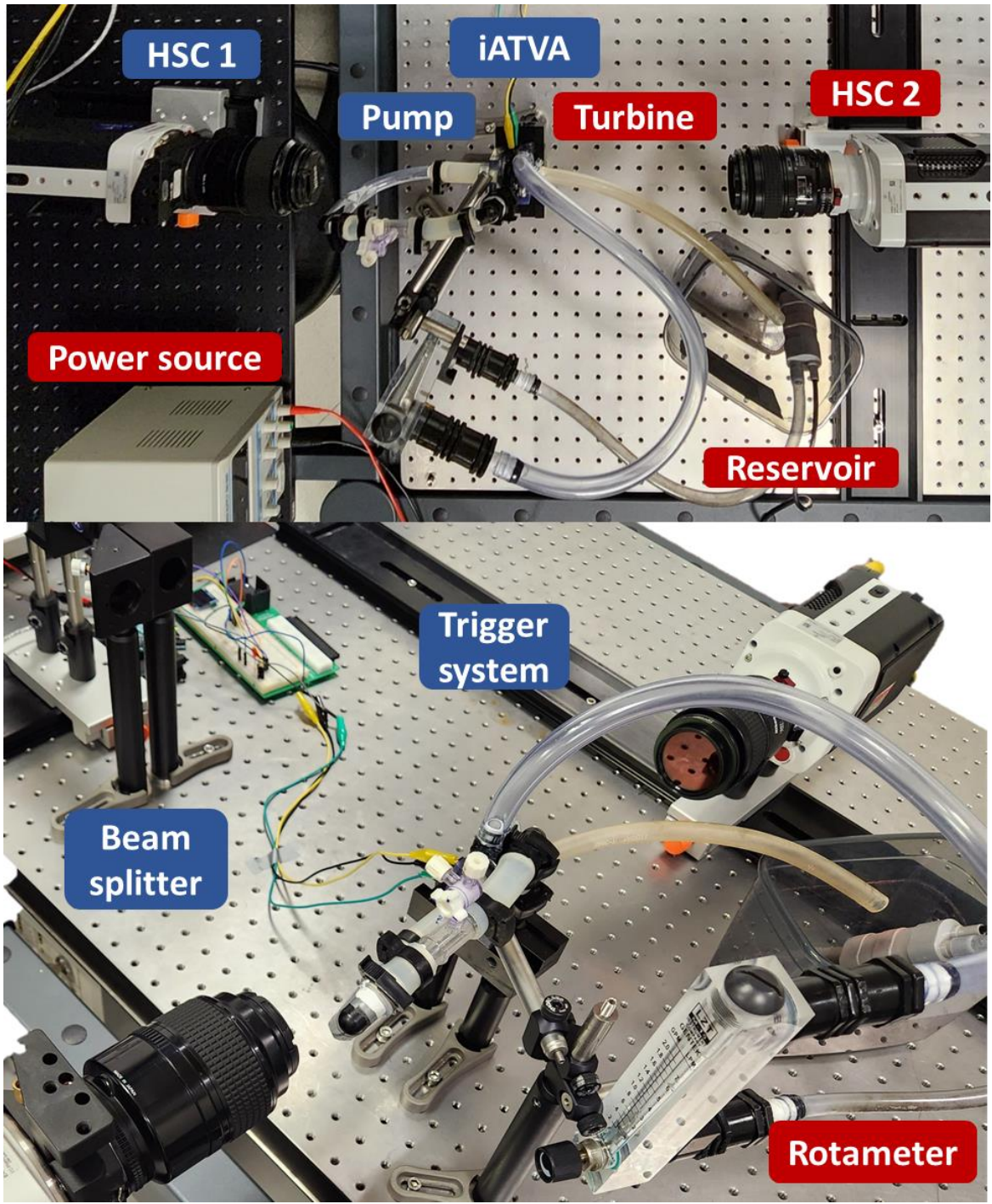

**Fig. 2** Selected views of the experiment set-up are provided. Two high speed cameras (HSC 1 and HSC 2), a hall-effect sensor-based trigger system controlled by an Arduino processor and the beam splitter are labeled. Other flow loop components include rotameter, reservoir, external pump, and power source for the external pump. Red and blue colors are used to highlight arterial and venous sections of iATVA, respectively.

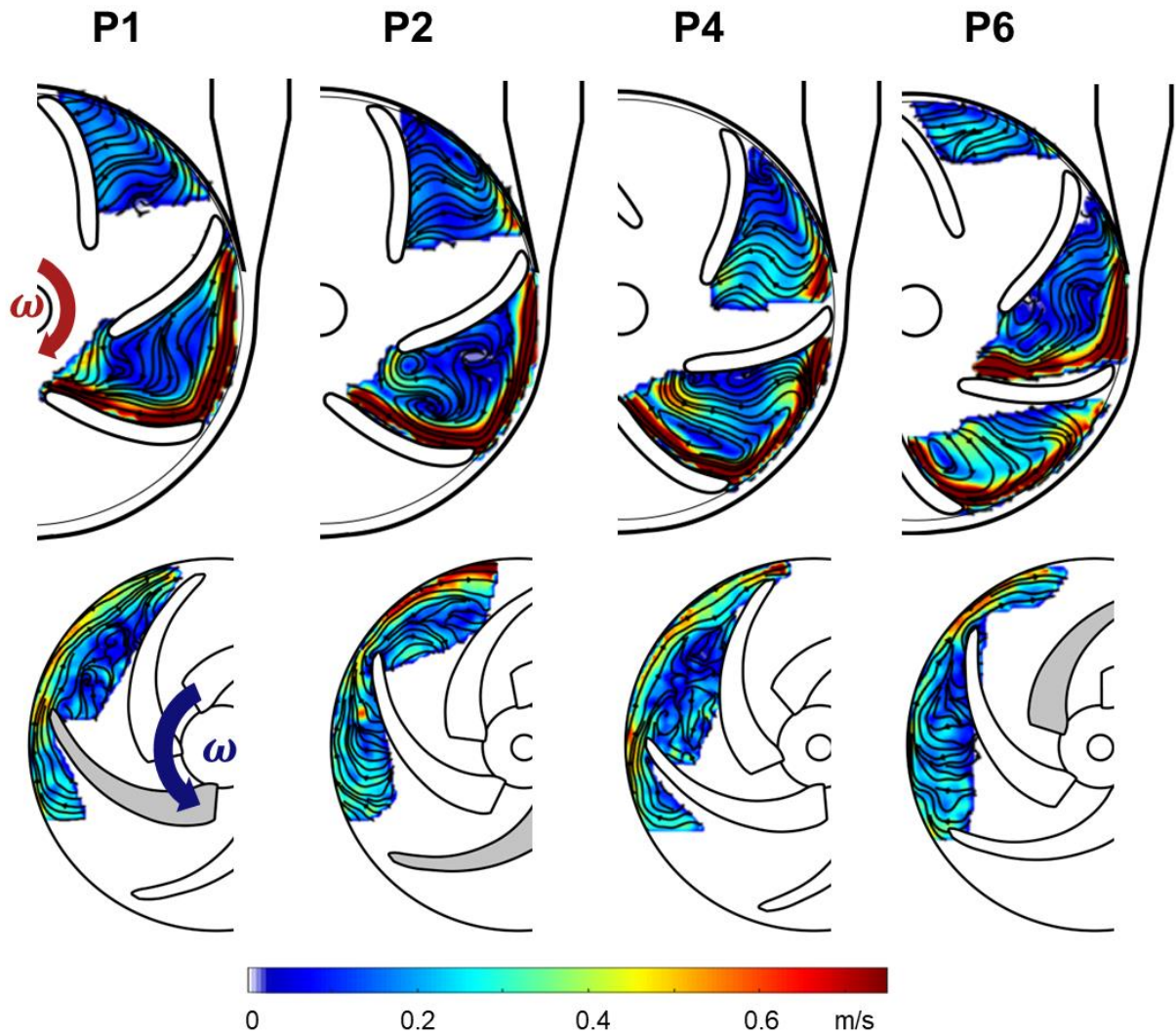

**Fig. 3** The simultaneous measurements of turbine and pump velocity fields at the second operating condition (Condition C2) are displayed. In total, seven phases were recorded for both impeller sides. However, since phases were mostly similar, only four of the phases are displayed for brevity. Averaging was employed with 5 phase-locked velocity fields to resolve the transient operation.

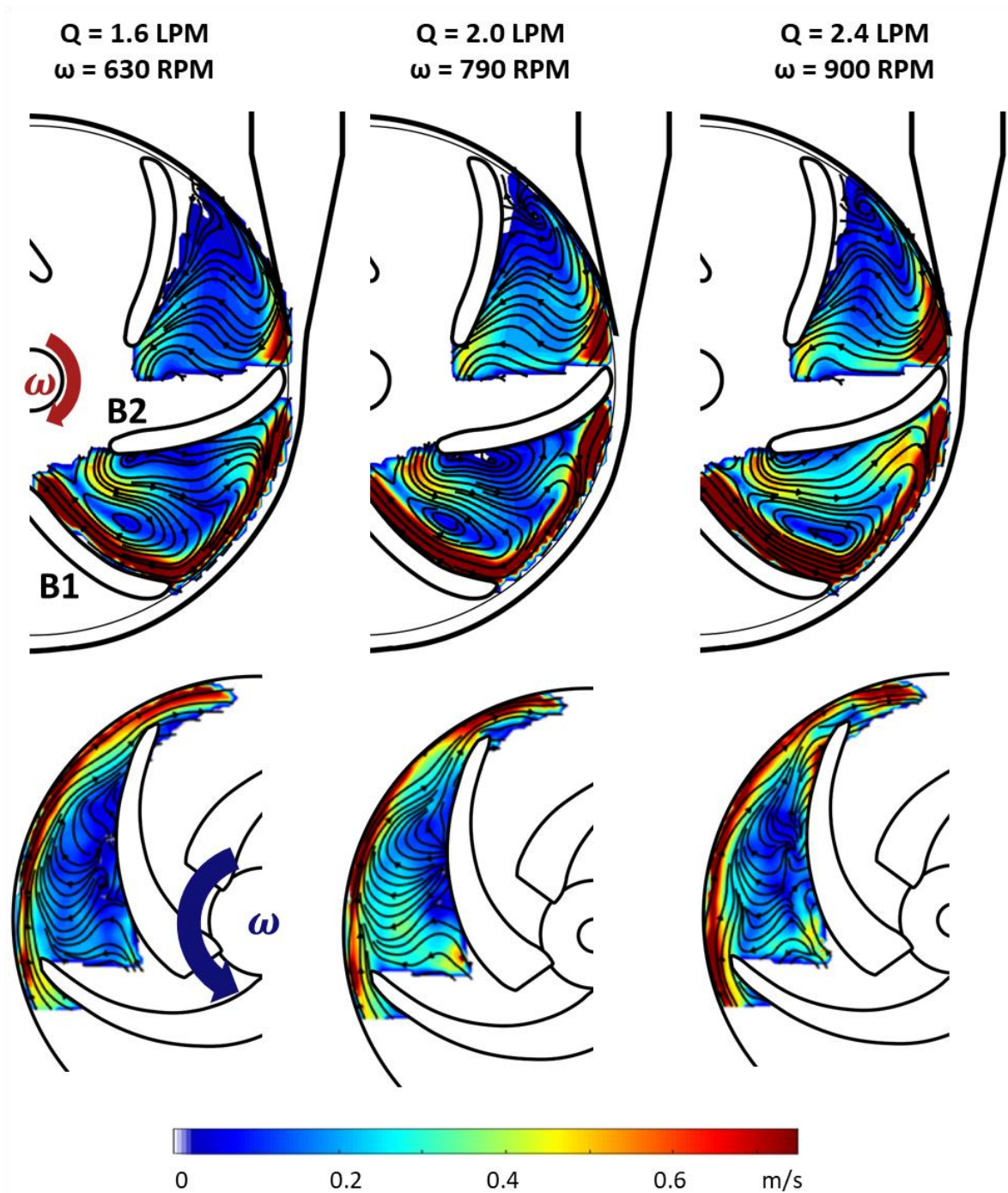

**Fig. 4** Relative velocity fields in turbine and pump impellers for three operating conditions (Conditions C1, C2 and C3). Operating conditions are determined according to the turbine flow rate ranging from 1.6 to 2.4 LPM. Turbine blades are labeled the same as in Fig. 3. Same color bar is used with Fig. 3 and for both turbine and pump sides. The mean jet flow velocity magnitudes in the pressure sides of B1 are 1.2, 1.5, 1.9 m/s for C1, C2, and C3 respectively.

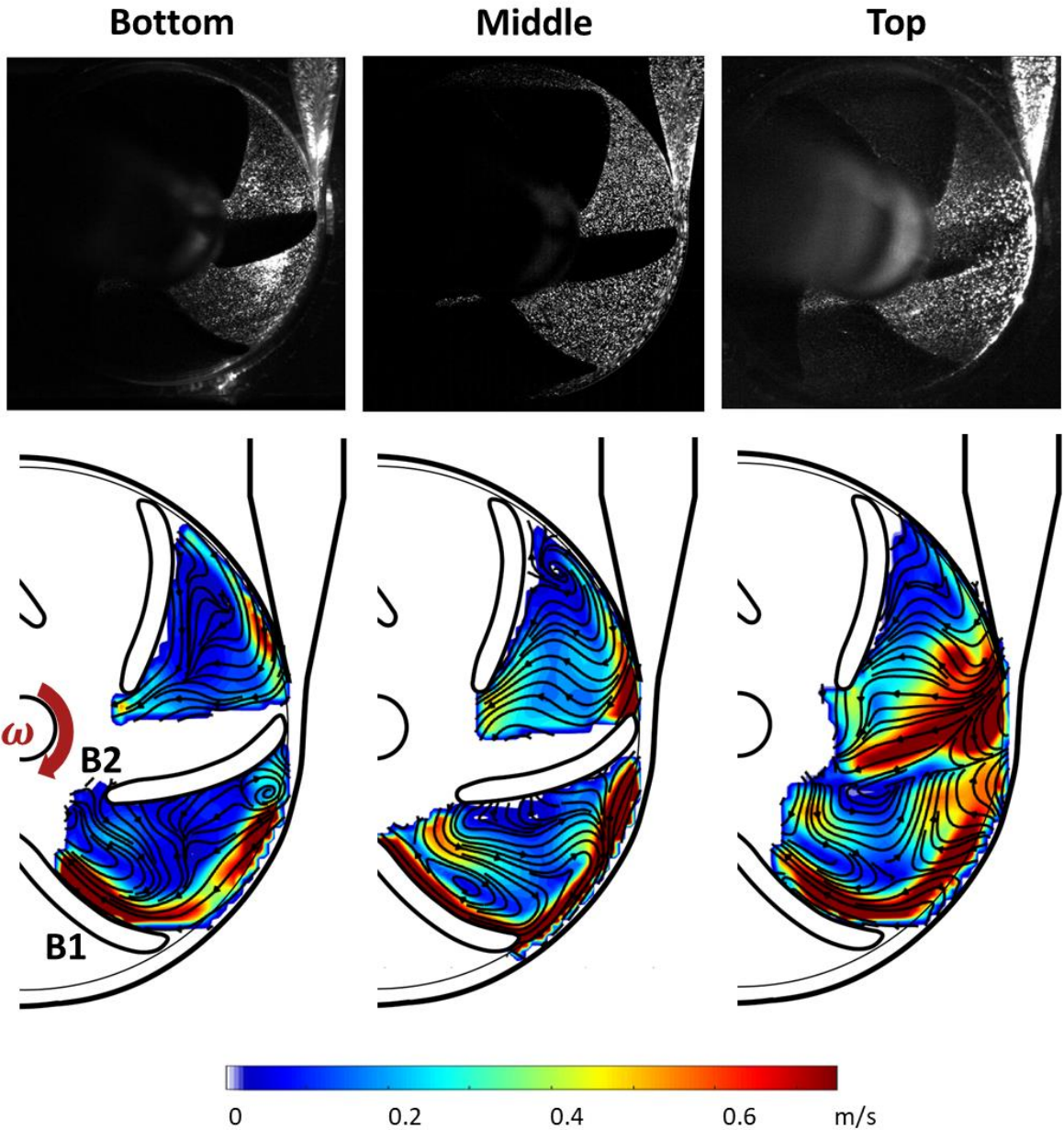

**Fig. 5** Velocity fields which are relative to the impeller rotation on the turbine impeller at three different sections (bottom, middle, and top) for the operating condition of turbine flow at 2 LPM. Raw images are presented in the first row and the corresponding velocity fields are presented in the second row. The blade aligned with the nozzle entrance is labeled as blade 2 (B2), and the next blade is labeled as B1. The leakage flow that passes through upon the gap in between the blade and the shroud is captured in the top section. The high speed of the jet flow dominates the velocity field, so the color bar is limited to 0.75 m/s to emphasize the flow structure in the blade passages. The mean jet flow velocity magnitudes at the pressure side of B1 are 1.3, 1.5, and 1.1. for each sections.

Page | 22

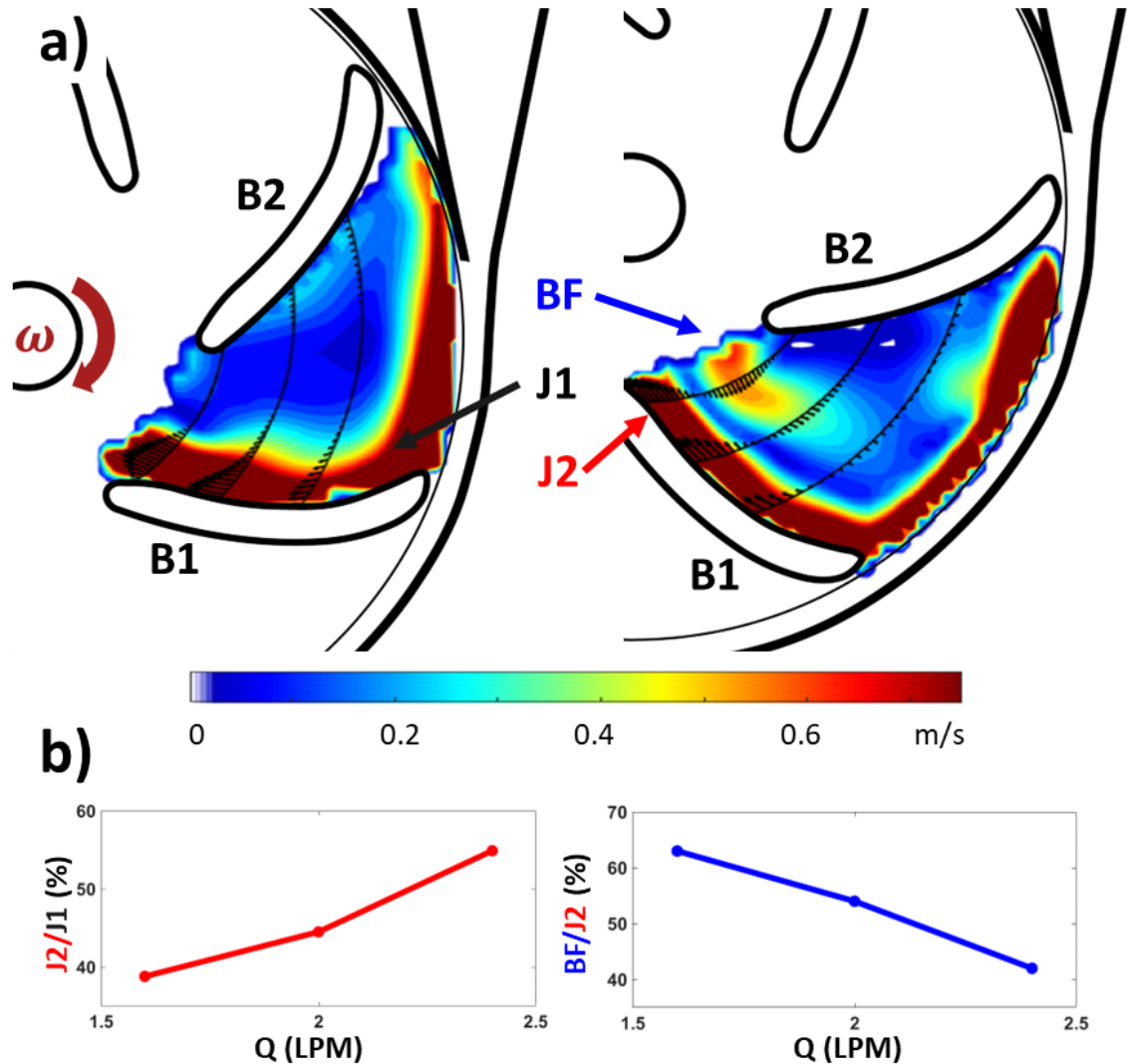

**Fig. 6** Quantitative visualization of velocity fields at two phases with velocity profiles. **a)** Most of the jet flow (J1) goes to the first blade passage in the left turbine phase, and the phase where the jet flow (J2) moves to the second blade passage is displayed in the right side. Backflow (BF) is marked in the right velocity field. **b)** The ratio of the velocity magnitudes of J2 to J1 for different operation conditions is displayed on the left side with an increasing trend. The ratio of the mean velocity magnitude of BF to J2 is presented on the right side with a decreasing trend. The mean BF velocity magnitudes were constant for different operation conditions.

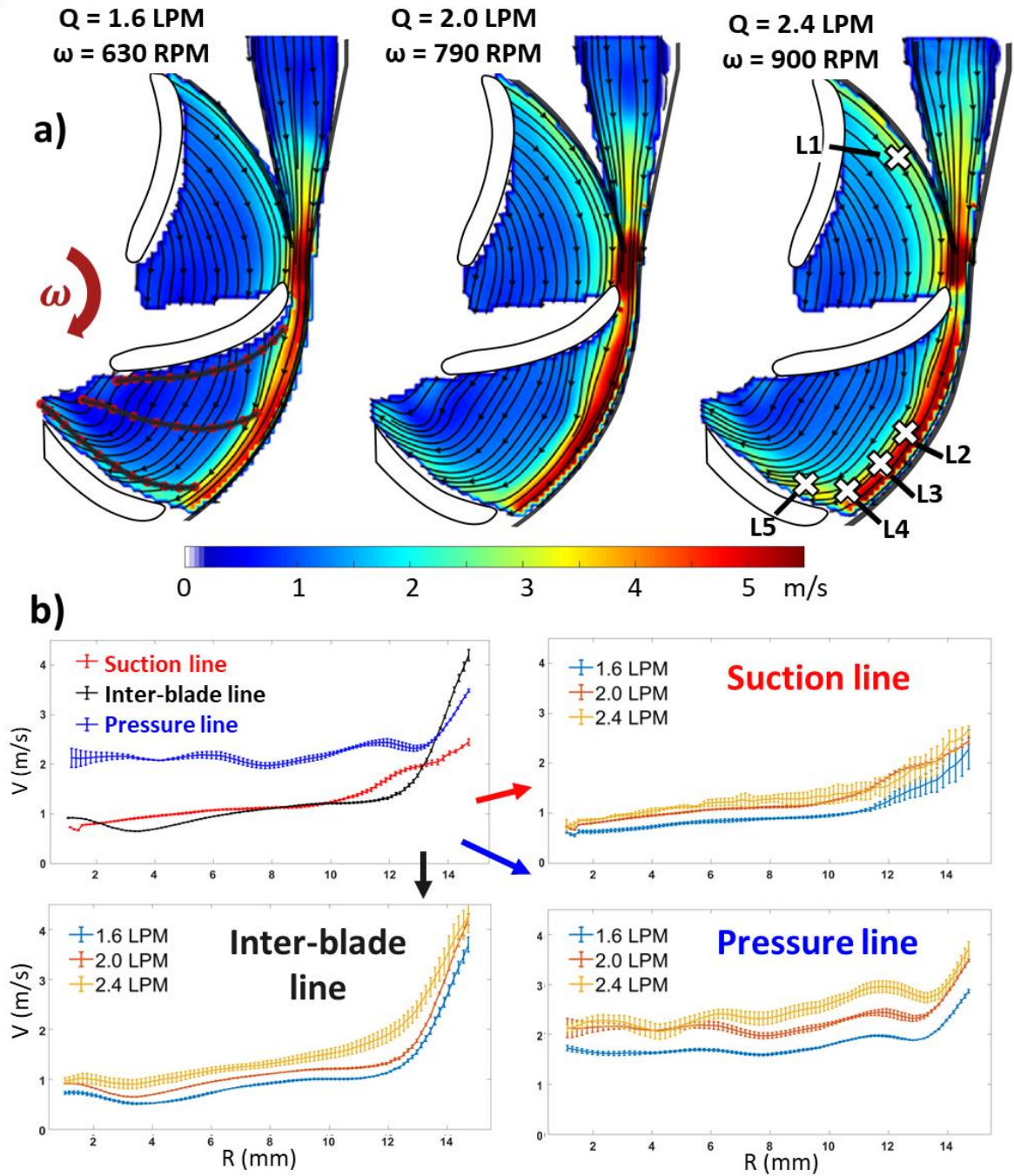

**Fig. 7** Velocity profiles along three different lines that are parallel to blade curvature. **a)** Absolute velocity contours for three different operating conditions. The suction, inter-blade and pressure lines are displayed in the first column. **b)** Velocity profiles along different sections and operating conditions. Radius is defined from the trailing edge to the leading edge (R). The top left plot is comparison of velocity fields for operating condition C2. Other plots are comparisons of the sections with three different operating conditions, C1, C2 and C3.

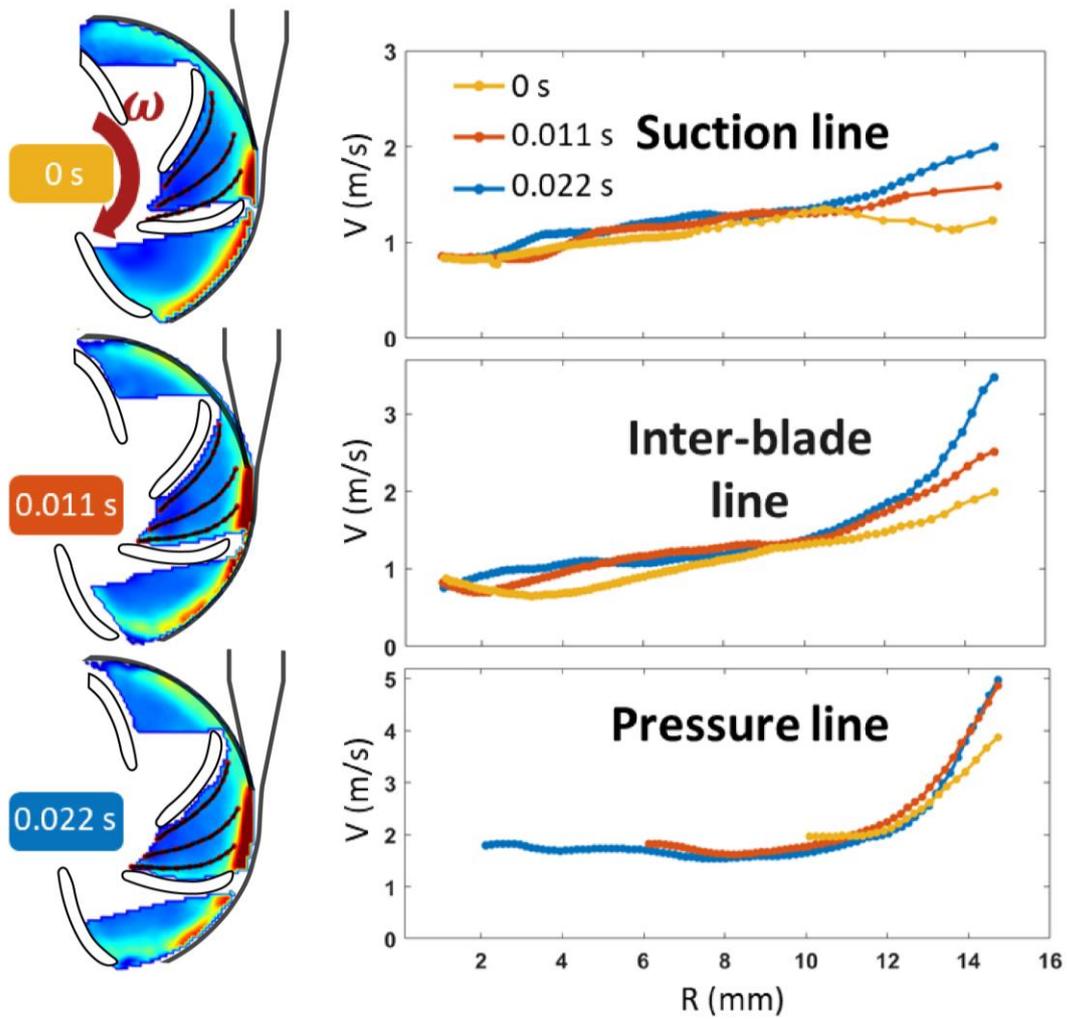

**Fig. 8** Velocity profiles along three different lines for three different phases (P1, P2, P3). Phases are displayed in the left side figure. Suction, inter-blade and pressure lines are displayed on the right side for each phases. Radius (R) is defined the same as in Fig. 7. Pressure sections had missing points near to the beginning of the radius due to the shadows, thus they were eliminated from the dataset.

# References

Atti, Varunsiri, Mahesh Anantha Narayanan, Brijesh Patel, Sudarshan Balla, Aleem Siddique, Scott Lundgren, and Poonam Velagapudi. 2022. "A Comprehensive Review of Mechanical Circulatory Support Devices." *Heart International* 16 (1): 37. https://doi.org/10.17925/HI.2022.16.1.37.

Blume, Elizabeth D., David N. Rosenthal, Joseph W. Rossano, J. Timothy Baldwin, Pirooz Eghtesady, David L.S. Morales, Ryan S. Cantor, et al. 2016. "Outcomes of Children Implanted with Ventricular Assist Devices in the United States: First Analysis of the Pediatric Interagency Registry for Mechanical Circulatory Support (PediMACS)." *The Journal of Heart and Lung Transplantation : The Official Publication of the International Society for Heart Transplantation* 35 (5): 578–84. https://doi.org/10.1016/J.HEALUN.2016.01.1227.

Bourque, Kevin, Christopher Cotter, Charles Dague, Daniel Harjes, Onur Dur, Julien Duhamel, Kaitlyn Spink, Kelly Walsh, and Edward Burke. 2016. "Design Rationale and Preclinical Evaluation of the HeartMate 3 Left Ventricular Assist System for Hemocompatibility." *ASAIO Journal (American Society for Artificial Internal Organs : 1992)* 62 (4): 375–83. https://doi.org/10.1097/MAT.0000000000000388.

Bowen, Robert E. S., Thomas J. Graetz, Daniel A. Emmert, and Michael S. Avidan. 2020. "Statistics of Heart Failure and Mechanical Circulatory Support in 2020." *Annals of Translational Medicine* 8 (13): 827–827. https://doi.org/10.21037/ATM-20-1127.

Corsini, Lucia, Valeria Dammicco, and James Moultrie. 2021. "Frugal Innovation in a Crisis: The Digital Fabrication Maker Response to COVID-19." *R and D Management* 51 (2): 195–210. https://doi.org/10.1111/RADM.12446.

Dunn, Jessica Lea, Erez Nusem, Karla Straker, Shaun Gregory, and Cara Wrigley. 2019. "Human Factors and User Experience Issues with Ventricular Assist Device Wearable Components: A Systematic Review." *Annals of Biomedical Engineering* 47 (12): 2431–88. https://doi.org/10.1007/S10439-019-02303-3/FIGURES/3.

Gil, Antonio, Roberto Navarro, Pedro Quintero, and Andrea Mares. 2023. "Hemocompatibility and Hemodynamic Comparison of Two Centrifugal LVADs: HVAD and HeartMate3." *Biomechanics and Modeling in Mechanobiology* 22 (3): 871–83. https://doi.org/10.1007/S10237-022-01686-Y/FIGURES/10.

Gorter, Thomas M., Dirk J. van Veldhuisen, Johann Bauersachs, Barry A. Borlaug, Jelena Celutkiene, Andrew J.S. Coats, Marisa G. Crespo-Leiro, et al. 2018. "Right Heart Dysfunction and Failure in Heart Failure with Preserved Ejection Fraction: Mechanisms and Management. Position Statement on Behalf of the Heart Failure Association of the European Society of Cardiology." *European Journal of Heart Failure* 20 (1): 16–37. https://doi.org/10.1002/EJHF.1029.

Hariharan, Prasanna, Kenneth I. Aycock, Martin Buesen, Steven W. Day, Bryan C. Good, Luke H. Herbertson, Ulrich Steinseifer, Keefe B. Manning, Brent A. Craven, and Richard A. Malinauskas. 2018. "Inter-Laboratory Characterization of the Velocity Field in the FDA Blood Pump Model Using Particle Image Velocimetry (PIV)." *Cardiovascular Engineering and Technology* 9 (4): 623–40. https://doi.org/10.1007/S13239-018-00378-Y/FIGURES/13.

Hashimoto, Shigehiro. 1989. "Erythrocyte Destruction under Periodically Fluctuating Shear Rate: Comparative Study with Constant Shear Rate." *Artificial Organs* 13 (5): 458–63. https://doi.org/10.1111/J.1525-1594.1989.TB01558.X.

Lipowsky, Herbert H., Shunichi Usami, and Shu Chien. 1980. "In Vivo Measurements of 'Apparent Viscosity' and Microvessel Hematocrit in the Mesentery of the Cat." *Microvascular Research* 19 (3): 297–319. https://doi.org/10.1016/0026-2862(80)90050-3.

Pekkan, Kerem, Ibrahim Basar Aka, Ece Tutsak, Erhan Ermek, Haldun Balim, Ismail Lazoglu, and Riza Turkoz. 2018. "In Vitro Validation of a Self-Driving Aortic-Turbine Venous-Assist Device for Fontan Patients." *The Journal of Thoracic and Cardiovascular Surgery* 156 (1): 292.

https://doi.org/10.1016/J.JTCVS.2018.02.088.

Ponnaluri, Sailahari V., Prasanna Hariharan, Luke H. Herbertson, Keefe B. Manning, Richard A. Malinauskas, and Brent A. Craven. 2023. "Results of the Interlaboratory Computational Fluid Dynamics Study of the FDA Benchmark Blood Pump." *Annals of Biomedical Engineering* 51 (1): 253–69. https://doi.org/10.1007/S10439-022-03105-W/TABLES/4.

Rowlands, Grant W., Bryan C. Good, Steven Deutsch, and Keefe B. Manning. 2018. "Characterizing the HeartMate II Left Ventricular Assist Device Outflow Using Particle Image Velocimetry." *Journal of Biomechanical Engineering* 140 (7). https://doi.org/10.1115/1.4039822/367296.

Sang, Xiaohu, and Xiaojun Zhou. 2017. "Investigation of Hydraulic Performance in an Axial-Flow Blood Pump with Different Guide Vane Outlet Angle." *Advances in Mechanical Engineering* 9 (8): 1–11. https://doi.org/10.1177/1687814017715423/ASSET/IMAGES/LARGE/10.1177_1687814017715423-FIG13.JPEG.

Sarkar, Soumodip, and Sara Mateus. 2022. "Doing More with Less - How Frugal Innovations Can Contribute to Improving Healthcare Systems." *Social Science & Medicine* 306 (August): 115127. https://doi.org/10.1016/J.SOCSCIMED.2022.115127.

Shu, Fangjun, Stijn Vandenberghe, Jaclyn Brackett, and James F. Antaki. 2015. "Classification of Unsteady Flow Patterns in a Rotodynamic Blood Pump: Introduction of Non-Dimensional Regime Map." *Cardiovascular Engineering and Technology* 6 (3): 230–41. https://doi.org/10.1007/S13239-015-0231-0.

Tamari, Y., K. Lee-Sensiba, E. F. Leonard, V. Parnell, and A. J. Tortolani. 1993. "The Effects of Pressure and Flow on Hemolysis Caused by Bio-Medicus Centrifugal Pumps and Roller Pumps: Guidelines for Choosing a Blood Pump." *The Journal of Thoracic and Cardiovascular Surgery* 106 (6): 997–1007. https://doi.org/10.1016/S0022-5223(19)33970-4.

Tsao, Connie W., Aaron W. Aday, Zaid I. Almarzooq, Cheryl A.M. Anderson, Pankaj Arora, Christy L. Avery, Carissa M. Baker-Smith, et al. 2023. "Heart Disease and Stroke Statistics-2023 Update: A Report From the American Heart Association." *Circulation* 147 (8): E93–621. https://doi.org/10.1161/CIR.0000000000001123.

Ucak, Kagan, Faruk Karatas, and Kerem Pekkan. 2024. "Effect of Impeller Phase on the FDA Blood Pump."

Valiathan, Marthanda Varma Sankaran. 2018. "Frugal Innovation in Cardiac Surgery Frugal Innovation: A Global Phenomenon." *Indian J Thorac Cardiovasc Surg* 34 (3). https://doi.org/10.1007/s12055-018-0652-0.

Wang, Shuai, Jianping Tan, Zheqin Yu, and Payman Jalali. 2020. "Shear Stress and Hemolysis Analysis of Blood Pump under Constant and Pulsation Speed Based on a Multiscale Coupling Model." *Mathematical Problems in Engineering* 2020. https://doi.org/10.1155/2020/8341827.

Wiegmann, L., S. Boës, D. de Zélicourt, B. Thamsen, M. Schmid Daners, M. Meboldt, and V. Kurtcuoglu. 2017. "Blood Pump Design Variations and Their Influence on Hydraulic Performance and Indicators of Hemocompatibility." *Annals of Biomedical Engineering* 46 (3): 417–28. https://doi.org/10.1007/S10439-017-1951-0/FIGURES/8.

Wiegmann, Lena, Bente Thamsen, Diane de Zélicourt, Marcus Granegger, Stefan Boës, Marianne Schmid Daners, Mirko Meboldt, and Vartan Kurtcuoglu. 2019. "Fluid Dynamics in the HeartMate 3: Influence of the Artificial Pulse Feature and Residual Cardiac Pulsation." *Artificial Organs* 43 (4): 363–76. https://doi.org/10.1111/aor.13346.